\documentclass[a4paper,fleqn]{cas-dc}
\usepackage{caption}
\usepackage[numbers]{natbib}
\usepackage[switch]{lineno}

\usepackage{color,xcolor}
\definecolor{amethyst}{rgb}{0.6, 0.4, 0.8}
\definecolor{alizarin}{rgb}{0.82, 0.1, 0.26}
\definecolor{green}{rgb}{0.55, 0.71, 0.0}
\definecolor{apricot}{rgb}{0.98, 0.81, 0.69}
\definecolor{auburn}{rgb}{0.43, 0.21, 0.1}
\definecolor{babyblueeyes}{rgb}{0.63, 0.79, 0.95}
\definecolor{bittersweet}{rgb}{1.0, 0.44, 0.37}
\definecolor{arsenic}{rgb}{0.23, 0.27, 0.29}

\usepackage{ulem}
\newcommand{\cfsout}{\bgroup\markoverwith{\textcolor{red}{\rule[0.5ex]{2pt}{0.4pt}}}\ULon}

\usepackage{hyperref}
\hypersetup{
    colorlinks=true,
    linkcolor=blue}

\begin{document}
\let\WriteBookmarks\relax

\shorttitle{Lorentz Invariance Violation in the BOAT}    

\shortauthors{F. Rescic et~al.}  

\title [mode = title]{Is There New Physics Beyond 30 TeV in the BOAT?}                      

\author[1,2]{Filip Rescic}[orcid=0000-0002-9664-5414]
\ead{filip.rescic@uniri.hr}

\author[3,4,5]{Luis Recabarren Vergara}[orcid=0009-0004-9449-8504]

\author[4]{Michele Doro}[orcid=0000-0001-9104-3214]
\author[1]{Tomislav Terzi\'c}[orcid=0000-0002-4209-3407]

\affiliation[1]{organization={University of Rijeka, Faculty of Physics},
            city={Rijeka},
            citysep={}, 
            postcode={51000}, 
            country={Croatia}}
\affiliation[2]{organization={Departamento de Fisica Teórica and Centro de Astropartículas y Física de Altas Energías (CAPA), Universidad de Zaragoza},
            city={Zaragoza},
            citysep={}, 
            postcode={50009}, 
            country={Spain}}
\affiliation[3]{organization={Centro di Ateneo di Studi e Attività Spaziali "Giuseppe Colombo"},
            addressline={Via Venezia 15}, 
            city={Padova},
            citysep={}, 
            postcode={ I-35131}, 
            country={Italy}}
\affiliation[4]{organization={Department of Physics and Astronomy, University of Padova},
            city={Padova},
            citysep={}, 
            postcode={I-35131}, 
            country={Italy}}
\affiliation[5]{organization={Istituto Nazionale Fisica Nucleare sez. Padova},
            city={Padova},
            citysep={}, 
            postcode={I-35131}, 
            country={Italy}}

\begin{abstract}
The exceptionally bright gamma-ray burst GRB~221009A, detected up to multi-TeV energies by the LHAASO observatory, provides a unique opportunity to probe possible deviations from standard photon propagation at extreme energies. In particular, it allows to test quadratic subluminal Lorentz Invariance Violation (LIV) scenarios through a potential enhancement of the observed flux at the highest energies. We argue that excesses in the GRB~221009A data currently classified as non-detections at energies $E \gtrsim 30$~TeV warrant further investigation, as they may indicate a recovery of the observable spectrum consistent with the LIV-induced suppression of $\gamma\gamma \to e^-e^+$ interactions during propagation. The absence of such a signature would allow one to exclude previously unexplored regions of the parameter space associated with the energy scale of new physics.

\end{abstract}

\begin{keywords}
Gamma-ray sources \sep Ground-based astronomy \sep Quantum gravity phenomenology
\end{keywords}

\maketitle

\section{Introduction}

Very-high-energy (VHE, $100\,\mathrm{GeV} < E < 100\,\mathrm{TeV}$) and ultra-high-energy (UHE, $E > 100\,\mathrm{TeV}$) gamma rays are produced in powerful non-thermal astrophysical sources such as gamma-ray bursts (GRBs) and active galactic nuclei (AGNs). During their propagation over cosmological distances, these photons are absorbed through $\gamma\gamma \to e^-e^+$ pair production interactions with the Extragalactic Background Light (EBL) \cite{Dominguez:2010bv,Saldana-Lopez:2020qzx} and the Cosmic Microwave Background (CMB)~\cite{nikishov1962,Gould:1967zza}.

Anomalies in photon absorption serve as a window to probing effects of Lorentz Invariance Violation (LIV), which provides one of the phenomenological frameworks for exploring possible imprints of Quantum Gravity at energies far below the Planck scale\footnote{$\mathrm{E}_{pl}\approx1.2\times 10^{19}$\,GeV.}~\cite{Addazi:2021xuf}. 
Previous theoretical studies have extended the Standard Model, exploring operators that break Lorentz Invariance explicitly. The first theoretical analyses of this sort were performed within an effective field theory approach~\cite{Colladay:1998fq, Myers:2003fd}. Given the interaction of interest in this work, we restrict our attention to the electrodynamic sector, focusing particularly on modifications in the photon sector. For a study exploring the phenomenological implications and experimental constraints of such modifications in the electron sector, see, e.g.,~\cite{Li:2022ugz}.
These modifications, arising from higher-dimensional operators suppressed by generic new physics scales $E_{\text{LIV},n}$, alter the photon dispersion relation, which takes the general form:
\begin{equation}
E^2 - \vec{k}^{\,2} = E^2 \sum_{n=1}^{\infty}S_n \left(\frac{E}{E_{\text{LIV},n}}\right)^n\,,
\label{eq:mdr_gen}
\end{equation}
where $E$ and $\vec{k}$ represent the energy and momentum of the photon, respectively. The sign factor $S_n=\pm 1$ introduces an ambiguity, corresponding to superluminal $(+)$ and subluminal $(-)$ scenarios. Phenomenological consequences of such modifications are typically studied on a term-by-term basis for specific orders $n$, with a focus on the linear ($n=1$) and quadratic ($n=2$) cases, as these yield the most significant effects and are more readily constrained by observations. For a general overview and phenomenological analysis of the different LIV modifications at the field-theoretical level, see the reviews~\cite{Jacobson:2005bg, mattingly2008tested,Liberati:2013xla}.

In general, LIV-induced modifications lead to a range of observable effects related to energy-dependent photon group velocity and altered interaction kinematics and dynamics.  
The corresponding bounds vary significantly between scenarios. A comprehensive overview of constraints and analysis techniques, particularly with data from Imaging Atmospheric Cherenkov Telescopes (IACTs), is provided in~\cite{Terzic:2021rlx}. 
Among other effects, the quadratic subluminal scenario introduces threshold modifications in the $\gamma\gamma\to e^-e^+$ process, giving rise to anomalous transparency at energies where standard physics predicts strong absorption. 
This manifests observationally as a spectral hardening or flux recovery at tens of TeV, precisely the energy range accessible to shower front detectors instruments such as LHAASO~\cite{LHAASO:2019qtb}. 

In this work, using the recently derived full $\gamma\gamma \to e^-e^+$ cross section in the quadratic subluminal LIV framework, which consistently accounts for modified kinematics and dynamics~\cite{Carmona:2024thn},
we investigate this scenario in the case of the extreme spectrum of GRB~221009A, as measured by LHAASO~\cite{LHAASO:2023kyg}. 
We propose that the high-energy end of the observed flux provides a particularly sensitive probe of quadratic subluminal LIV. We show that the unexplored energy range above 30\,TeV in GRB~221009A may contain the characteristic flux recovery expected from quadratic subluminal LIV, and we motivate a targeted search in this energy range. In addition, we discuss the benchmark sensitivities to these effects, of future gamma-ray observatories: the Southern Wide-field Gamma-ray Observatory (SWGO)~\cite{SWGO:2025taj} and the Cherenkov Telescope Array Observatory (CTAO)~\cite{CTAConsortium:2017dvg}, showing their potential for flux anomaly observability in sources such as the blazar PKS~2155-304. SWGO is a proposed next-generation wide-field gamma-ray observatory in the Southern Hemisphere, designed to provide continuous monitoring of the sky at multi-TeV energies. On the other hand, CTAO, with installations in both the Northern and Southern Hemispheres, is a proposed future IACT facility optimized for pointed observations up to hundreds of TeV, complementary to the wide-field, high-duty-cycle capabilities of SWGO.

This paper is organized as follows: in Section~2 we introduce the phenomenology of photon absorption in the quadratic subluminal LIV scenario; in Section~3 we describe the methodology used to model gamma-ray attenuation; in Section~4 we present our results for GRB~221009A; finally, Section~5 summarizes our conclusions and discusses future prospects.

\section{Phenomenology}

In quadratic subluminal LIV, we rewrite~\autoref{eq:mdr_gen} as
\begin{equation} 
E^2 - \vec{k}^2 = - \frac{E^4}{\Lambda^2}\,,
\label{eq:mdr}
\end{equation} 
where $\Lambda\equiv E_{\text{LIV},2}$. As a consequence, ~\autoref{eq:mdr} changes both the reaction threshold and the cross section for pair production. The corresponding threshold condition is 
\begin{equation}
2E\varepsilon(1-\cos\theta) - \frac{E^4}{\Lambda^2} \geq 4m_e^2\,,
\label{eq:thr}
\end{equation}
where $m_e$ is the electron mass, $\varepsilon$ is the EBL or CMB background photon energy and $\theta$ is the angle of incidence. Although the correction term $(E/\Lambda)^2$ is minuscule even for PeV gamma rays, given the current constraint on $\Lambda$ (see below), the $E^4/\Lambda^2$ contribution can still produce an observable effect at energies far below~$\Lambda$. The reason is the appearance of a $\Lambda$-dependent upper threshold  making the interaction take place only within a finite range of energies (see the discussion in Section~II~B of~\cite{Carmona:2024thn} for more details). For fixed values of the kinematical variables $\varepsilon$ and $\theta$, this interval is given by
\begin{equation}
    \label{eq:range}
    \frac{2m_{e}^{2}}{\varepsilon(1-\cos{\theta})}\leq E\leq[2\varepsilon(1-\cos{\theta})\Lambda^{2}]^{1/3}\,.
\end{equation}
This is in contrast to the case of Special Relativity (SR)~\cite{PhysRev.46.1087}, where once the first threshold is reached, the interaction can occur for all energies of the gamma ray beyond it. 

In SR, the cross section for the pair production process $\gamma\gamma\to e^-e^+$ was calculated long ago and is now referred to as the Breit-Wheeler (BW) cross section~\cite{PhysRev.46.1087}. 
In the context of LIV, however, determining the cross section of this process has been an open issue. Previous works used approximations based on the BW result~\cite{Martinez-Huerta:2020cut, Blanch:2001hu,HESS:2019rhe,Tavecchio:2015rfa,Fairbairn:2014kda,Abdalla:2018sxi}. 
In~\cite{Carmona:2024thn} limitations of these approaches were discussed and a more accurate expression for the cross section was proposed, which we adopt here.

\subsection{GRB~221009A}

When searching for LIV, we are primarily interested in \textit{hard and bright} gamma-ray sources, as their spectra extend to the highest energies with sufficient photon statistics to place meaningful constraints. The interaction of VHE and UHE gamma rays, produced in powerful astrophysical accelerators, with low-energy photons from EBL or CMB provides a natural framework to probe potential LIV effects. Among the known sources capable of generating such energetic emission, gamma-ray bursts (GRBs) stand out. 
The highly non-thermal spectra of GRBs, extending up to the TeV regime, together with their cosmological distances and extreme luminosities, make them exceptional laboratories for testing fundamental physics, including possible LIV-induced effects. 

Several GRBs have been detected across the GeV--TeV energy band~\cite{MAGIC:2019lau, Abdalla:2019dlr}, including at least five events with firm detections of TeV afterglow emission~\cite{Foffano:2025bhj} and several additional low-significance candidates~\cite{Miceli:2022efx}. The exceptionally bright GRB~221009A also known as the \textit{brightest of all time} (BOAT), demonstrates that GRBs can accelerate particles to extreme energies and emit photons well into the TeV regime. 

The BOAT (redshift $z=0.151$), detected by LHAASO, represents an optimal candidate for exploring potential LIV-induced spectral signatures in the TeV energy range. The BOAT, with an isotropic-equivalent energy release of $E_{\gamma, \mathrm{iso}} \sim 10^{54}$~erg, was first detected on October 9, 2022, by the \textit{Fermi} Gamma-ray Burst Monitor~\cite{Lesage:2023vvj} and the \textit{Swift} Burst Alert Telescope~\cite{Williams:2023sfk}, which subsequently triggered follow-up observations by LHAASO. The LHAASO Water Cherenkov Detector Array (WCDA) recorded more than $6.4\times10^4$ photons in the $[200~\mathrm{GeV},7~\mathrm{TeV}]$ range, with a statistical significance exceeding $250\sigma$ \cite{LHAASO:2023kyg}. Additionally, the LHAASO-KM2A detector registered photons up to $\sim13~\mathrm{TeV}$ \cite{LHAASA:2023pay}. The entire event lasted about $6000$~s within LHAASO’s field of view. Furthermore, $4536$~s after the initial trigger, the Carpet-2 collaboration reported the detection of a single photon-like event with an energy of $E=251$~TeV coming from the BOAT direction~\cite{Dzhappuev2022}. This event has been recently reanalyzed by the Carpet-3 collaboration obtaining a reconstructed energy of $E=300$~TeV~\cite{Carpet-3Group:2025fcs}.

\subsection{Previous LIV studies with the BOAT}

A number of recent studies have explored the potential of the BOAT for testing Lorentz invariance. Time-of-flight analyses, based on searches for energy-dependent photon group velocities, were performed in~\cite{Piran:2023xfg, Xi:2025ruv} within both linear and quadratic LIV scenarios.

Regarding anomalous transparency studies, analyses within the linear LIV scenario such as~\cite{LHAASO:2024lub, Li:2023rhj, Li:2023rgc, Li:2022wxc, Li:2022vgq} typically lead to limits where the LIV scale approaches or exceeds the Planck scale. More recently, works incorporating the Carpet-3 measurement into their fits have been carried out in the quadratic subluminal scenario. We refer the reader to~\cite{Ofengeim:2025jsw} and~\cite{Satunin:2025hbk} (and references therein) for a detailed discussion. These analyses infer limits of order $\Lambda \sim 10^{-7} \mathrm{E}_{pl}$, showing that in this scenario there remains a phenomenological window of interest, which motivates the present study. We emphasize, however, that given the non-conclusive nature of the Carpet-3 observation, our analysis relies exclusively on the robust GRB~221009A LHAASO dataset. 

We adopt $10^{-7}\mathrm{E}_{pl}$ as a starting order–of–magnitude benchmark for our analysis, as it is consistent with the scales suggested by single and multi–source studies (see also~\cite{HESS:2019rhe, Lang:2018yog}).

\section{Methodology}

The differential photon flux observed at Earth at energy $E$, and the intrinsic flux emitted by a source at redshift $z_s$ are related through 
\begin{equation}
    \frac{dN_\mathrm{obs}}{dE} = \frac{dN_{\mathrm{int}}}{dE}\times e^{-\tau(E,z_s; \Lambda)}\,.
    \label{eq:absspec}
\end{equation}
The attenuation factor represents the survival probability of gamma rays, with the optical depth $\tau(E, z_s; \Lambda)$ quantifying the transparency of the Universe. The optical depth depends on the photon-photon cross section $\sigma$, the low-energy background photon density $n$, the incidence angle $\theta$, and the cosmological effects on photon propagation across the Universe. This can be written as
\begin{align}
    \tau(E,z_s&; \Lambda) = \int_{0}^{z_s}dz\,\frac{dl}{dz}\int_{-1}^{1} d\cos\theta \left(\frac{1-\cos\theta}{2}\right) \notag \\ \times &\int_{\varepsilon_\mathrm{thr}(E,\theta,z; \Lambda)}^\infty d\varepsilon \; n(\varepsilon,z) \,\sigma(E(1+z),\varepsilon,\theta; \Lambda) \,.
    \label{eq:opa}
\end{align}
Here, $\varepsilon_{\mathrm{thr}}(E,\theta,z; \Lambda)$ is the minimum value of the background photon energy necessary to satisfy the threshold condition~\autoref{eq:thr}. The standard SR expressions are recovered in the limit $\Lambda \to \infty$. The term $dl/dz$ describes how the traversed path of the photon depends on the redshift. In the $\Lambda$CDM cosmological model, it is given by
\begin{equation}
    \frac{dl}{dz}=\frac{1}{(1+z)H_0\sqrt{\Omega_m(1+z)^3+\Omega_\Lambda}}\,.
\end{equation}
Following~\cite{Dominguez:2010bv, Saldana-Lopez:2020qzx}, we adopt $\Omega_m =0.3$, $\Omega_{\Lambda}=0.7$ and $H_0= 70$  km s$^{-1}$ Mpc$^{-1}$ for the matter density, the vacuum energy density, and the present value of the Hubble constant, respectively. 

We extrapolate the  intrinsic spectrum  of GRB~221009A in its highest state, corresponding to the time interval from 5 to 14~s after the start of LHAASO afterglow observations, into the $100$~TeV energy range and investigate the effect of gamma-ray absorption on EBL and CMB assuming different values for the energy scale $\Lambda$. Then we identify the energy intervals in which LIV-enabled flux recovery could be detected by investigating the signal content of the energy bins above $\sim30$\,TeV.

\section{Results}

In~\autoref{fig:boat_liv}, we present the intrinsic spectrum of GRB 221009A. It was obtained following the procedure from~\cite{Miceli:2024yhv}, where the authors took intrinsic spectrum in the 5 reported time bins and averaged it over time. Given the maximum energy of 13\,TeV reported in~\cite{LHAASA:2023pay}, the final intrinsic spectrum was modeled as a power law with exponential cutoff (ECPL) defined as
\begin{equation}
    \frac{dN_{\mathrm{int}}}{dE} = N_0 \left(\frac{E}{E_0}\right)^{-\alpha} \exp{\left(-E/E_\mathrm{cut}\right)}\,,
    \label{eq:int_ecpl}
\end{equation}
with $E_0 = 1$\,TeV, $\alpha=2.3$, and $E_\mathrm{cut} = 13$\,TeV. In our case, we used the highest BOAT state as for the flux normalization, $N_0 = 3.20\times 10^{-6}\,$TeV$^{-1}$\,cm$^{-2}$\,s$^{-1}$. The figure also shows the spectra attenuated by the background composed of EBL and CMB, obtained by multiplying the intrinsic spectrum by the survival probability (\autoref{eq:absspec}). We employ the EBL model of Saldana-Lopez et al. (2021)~\cite{Saldana-Lopez:2020qzx}. The attenuated spectra are labeled with the corresponding energy scales $\Lambda$, and SR, which represents the standard scenario. In particular, the light-blue dashed curves show the spectra in the quadratic subluminal LIV scenario (\autoref{eq:mdr}), for six energy scales $\Lambda = [1.0,\,1.3,\,2.0,\,2.6,\,5.0,\,9.0]\times10^{-7}\mathrm{E}_{pl}$.

\begin{center}
\includegraphics[width=\linewidth]{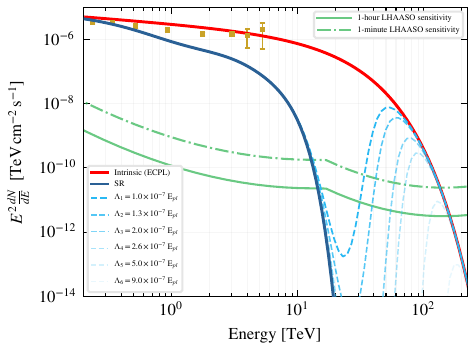}
\captionof{figure}{Intrinsic average data points of GRB~221009A (yellow markers) corresponding to the highest-emission state in the time interval between 5 and 14 seconds after the start of afterglow observations, extracted from the supplementary material of~\cite{LHAASO:2023kyg}, together with the corresponding intrinsic spectrum~(\autoref{eq:int_ecpl}) obtained from the time-averaged fit~\cite{Miceli:2024yhv} and normalized to the corresponding highest-emission interval. The solid blue line shows absorbed spectrum in SR scenario assuming EBL+CMB background. The dashed light-blue curves represent spectra modified by LIV effects for six $\Lambda$ values, using the $\gamma\gamma \rightarrow e^{-}e^{+}$ modified cross section from~\cite{Carmona:2024thn}. The EBL model used is that of Saldana-Lopez et al.~\cite{Saldana-Lopez:2020qzx}. The green curves indicate the LHAASO sensitivity scaled to observation times of 1~hour, and 1~minute.}
\label{fig:boat_liv} 
\end{center}

The attenuated spectra are compared with the 1-minute and 1-hour LHAASO sensitivities, which were estimated by multiplying the 1-year LHAASO sensitivity~\cite{Vernetto:2016gro} by a factor $\sqrt{1\,\mathrm{year} / T_\mathrm{obs}}$, 
where $T_\mathrm{obs}$ stands for 1 minute and 1 hour, respectively. Assuming that the intrinsic spectrum of the BOAT is indeed extended to energies $E \gtrsim 30$~TeV, it is evident that LHAASO is able to detect excess flux if LIV-reduced gamma-ray absorption were present. Therefore, we strongly advocate for a dedicated analysis of signal in the energy bins above 30\,TeV. 

In~\autoref{appendix:dom_modeling}, we show the LIV-modified spectra obtained using the Dom\'inguez et al. (2011)~\cite{Dominguez:2010bv} and Finke et al. (2022)~\cite{Finke:2022uvv} EBL models and discuss the resulting differences. While moderate deviations among the models appear in the strongly absorbed energy range around the spectral minimum, the high-energy LIV-induced recovery region remains essentially unchanged. This is expected, since our analysis focuses on the energy range in which the absorption becomes strongly suppressed due to LIV kinematics. Consequently, the position of the spectral recovery provides a robust probe of the LIV scale, independent of the specific EBL model employed. In this sense, the observation of the LIV-modified SED maximum lifts the degeneracy between the choice of EBL model and the inferred value of~$\Lambda$.

In addition, we find a relation between the energy, denoted by $E^*$, at which the measured LIV-modified spectral energy distribution (SED) would reach its maximum value (before approaching the intrinsic spectrum), and the corresponding value of the scale $\Lambda$. \cite{Jacob:2008gj} (see Eq. 4) derives a formula for calculating the gamma-ray energy at which the reaction energy threshold, in terms of the background photon energy, is minimal given a certain LIV energy scale. This roughly corresponds to the gamma-ray energy at which the gamma-ray absorption is the strongest, marking the onset of LIV effects. However, the difference between the expected fluxes in SR and LIV at those energies are difficult to differentiate past the measurement uncertainties. Instead, the gamma-ray flux recovery as the one expected from LIV, visible at $E^*$ that we propose, is not predicted by known emission models in SR, and is therefore a clear indication of LIV. 

As can be seen in~\autoref{eq:range}, the upper threshold scales as $\Lambda^{2/3}$. To test this dependence, we fit the relation between the energy at which the LIV-modified spectral maxima occur, $E^*$, and $\Lambda$ for the six values considered, assuming a functional dependence $E^{*}(\Lambda)=a\,\Lambda^{2/3}+b$. The maxima were determined by interpolating the LIV-modified SED curves. To assess the robustness of this relation, we repeated the analysis assuming two alternative intrinsic spectral models: a power-law (PL) and a log-parabola (LP). The PL model is given by
\begin{equation}
    \frac{dN_{\mathrm{int}}}{dE} = N_0 \left(\frac{E}{E_0}\right)^{-\alpha}\,,
    \label{eq:int_pwl}
\end{equation}
while the LP model is obtained by adding a curvature term $\beta$ in the spectral index:
\begin{equation*}
    \text{LP: } \alpha \rightarrow \alpha + \beta \log\!\left(\frac{E}{E_0}\right)\,.
\end{equation*}
For the LP model, the parameters $\alpha$ and $\beta$ are obtained by fitting the highest-state data points (see~\autoref{fig:boat_liv}), yielding $\alpha = 2.56$ and $\beta = 1.21$. For the PL model, the value of $\alpha$ is kept identical to the ECPL case. In both cases, the parameters $N_0$ and $E_0$ are also kept identical to those of the ECPL case (see below~\autoref{eq:int_ecpl}).

This is illustrated in~\autoref{fig:linear_relation}, where a linear fit assuming the scaling $E^*\propto \Lambda^{2/3}$ reproduces the functional dependence consistently. The best-fit values and uncertainties of the parameters $a$ and $b$ are reported in~\autoref{tab:fitparameters}. It is worth remarking that this relation provides a simple law for estimating accessible new-physics energy scales, given the sensitivity of a gamma-ray observatory.

We note that harder intrinsic spectra naturally enhance the observability of LIV-induced transparency effects. In particular, the delayed suppression of the intrinsic flux allows the LIV spectral recovery to remain above the experimental sensitivity up to significantly higher energies.

For the extreme PL case, assuming the spectrum persists to sufficiently high energies, the LIV-induced recovery could remain observable even near the PeV regime. Given a detector sensitive in that energy range, this would allow one to probe LIV scales several orders of magnitude higher than those considered in the present work, and beyond the range currently explored in the literature using extragalactic gamma-ray absorption studies.

\begin{center}
\includegraphics[width=\linewidth]{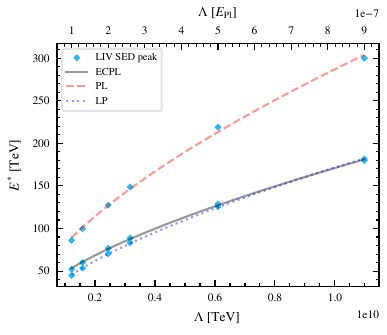}
\captionof{figure}{Relation between the LIV SED peak energy $E^{*}$ (light blue diamond markers) and the corresponding six values of the LIV energy scale $\Lambda$. Results are shown for the three different choices of the intrinsic BOAT emission spectrum: ECPL corresponding to the model used in \autoref{fig:boat_liv}, PL obtained by turning off the exponential term, and a LP model. For the LP case, $\alpha = 2.56$ and $\beta = 1.21$ are obtained from a fit to the highest-state data points. For all spectral models considered, the extracted peak energies $E^*$ follow the expected functional dependence, $E^*\propto\Lambda^{2/3}$. This corresponds to the behaviour of the upper threshold (\autoref{eq:range}). The best-fit parameters values are shown in~\autoref{tab:fitparameters}.}
\label{fig:linear_relation}
\end{center}

\begin{table}[H!]
\caption{Best-fit values of the parameters $a$ and $b$ for the relation $E^{*}(\Lambda) = a\Lambda^{2/3} + b$, obtained for the three intrinsic spectral models shown in~\autoref{fig:linear_relation}. The corresponding uncertainties are $\sigma_a=3.1\times10^{-7}$  and $\sigma_b = 8.7\times10^{-1}$, and are identical in all three cases.}
\label{tab:fitparameters}
\centering
\begin{tabular}{l c c }
\toprule
Spectral model & $a$ & $b$\\
\midrule
ECPL & $3.4\times10^{-5}$ & $1.5\times10^{1}$ \\
PL & $5.7\times10^{-5}$ & $2.4\times10^{1}$ \\
LP & $3.6\times10^{-5}$ & $4.8$ \\
\bottomrule
\end{tabular}
\end{table}

\section{Conclusion}

In this work, we highlight the potential of TeV gamma rays from GRB~221009A, detected by LHAASO, to constrain the influence of LIV on the spectrum in the quadratic subluminal scenario for the process $\gamma\gamma \to e^-e^+$. To this end, we use the recently derived modified cross section~\cite{Carmona:2024thn}. Having no results of the LHAASO data analysis in these energy bins (signal or upper limits), we were not able to set constraints on the LIV energy scale. However, the results of our analysis motivate further investigation of apparent excesses with insufficient statistical significance, i.e., non-detections, in $E \gtrsim 30$~TeV regime of the BOAT data, to search for spectral anomalies consistent with LIV or to place more stringent lower bounds on the scale~$\Lambda$. We verified that our conclusions are robust against the choice of EBL model, obtaining
consistent phenomenological behavior when using the Saldana-Lopez et al. (2021), Dom\'inguez et al. (2011) and Finke et al. (2022) models.

We also emphasize the quasi-empirical relation found between the SED maxima in the LIV cases and the corresponding scale $\Lambda$. In particular, we verified that the energy at which the LIV-modified SED maximum occurs scales proportionally to $\Lambda^{2/3}$. This behavior has no dependence on the EBL model or the intrinsic spectrum of the source for the energy scales $\Lambda$ considered. Moreover, this relation provides a direct connection between observable spectral features and the underlying scale of new physics. This scaling relation allows one to estimate, from the maximum energy accessible to a given experiment, the range of $\Lambda$ values that can be effectively probed within the quadratic subluminal scenario. In this context, the advent of currently operating and future wide–field, high-duty-cycle detectors such as LHAASO and SWGO, together with next-generation IACT facilities such as CTAO, significantly enhances the chances of capturing high-energy transients and enabling increasingly precise tests of LIV-induced spectral anomalies. 

More generally, the sensitivity of gamma-ray observations to LIV effects is controlled also by the propagation distance and by the dominant background photon field in the relevant energy range. In particular, observations of galactic sources extending into the PeV regime, where the CMB provides the dominant target photon field, can probe LIV energy scales several orders of magnitude above those accessible with TeV observations of extragalactic sources. This highlights the importance of accounting for the interplay between the gamma-ray energy and the propagation distance when discussing searches for LIV. In this respect, dedicated studies of galactic UHE sources observed by LHAASO~\cite{LHAASO:2021gok,LHAASO:2024psv} are expected to play a crucial role in significantly extending the region of the new-physics parameter space that can be tested.

\section*{Acknowledgments}
The authors would like to acknowledge networking support by COST Actions CA18108 QGMM (Quantum Gravity in the Multi-Messenger era, \url{https://qg-mm.unizar.es/}) and CA23130 BridgeQG (Bridging high and low energies in search of Quantum Gravity, \url{https://www.cost.eu/actions/CA23130/}.
M.D. acknowledges support from the Italian MUR Departments of Excellence grant 2023-2027 “Quantum Frontiers”. F.R. acknowledges the support from the Croatian Science Foundation (HrZZ) Outbound Mobility grant MOBDOK-2023. 
T.T. and F.R. acknowledge the support from the University of Rijeka through projects uniri-iskusni-prirod-23-24 and uniri-iz-25-119, and from the Croatian Science Foundation (HrZZ) Project IP-2022-10-4595. We thank the anonymous referees for useful comments on this work.   

\paragraph{Data Availability Statement}
All materials necessary to reproduce the figures in this work are available in the repository~\cite{rescic_2026_20345064}.

\appendix

\section{Testing different EBL models}\label{appendix:dom_modeling}

We tested the LIV-modified BOAT spectrum using the Dom\'inguez et al. (2011)~\cite{Dominguez:2010bv} and Finke et al. (2022; Model A)~\cite{Finke:2022uvv} EBL models. We find no significant changes in the spectrum in the low-absorption region at energies $E \gtrsim 30$~TeV, where the LIV-induced recovery takes place. 

In contrast, in the energy range characterized by stronger absorption, $E \lesssim 30$~TeV, corresponding to the region around the spectral minimum, moderate differences between the models are observed, as shown in~\autoref{fig:boatdominguez}. In particular, the onset of the LIV-induced deviations from the SR prediction depends on the adopted EBL model. This behavior can be traced back to the different infrared photon densities predicted by the corresponding EBL spectra (see, e.g., Fig.~5 of~\cite{Finke:2022uvv}), which determine the absorption strength in the relevant energy range.

While moderate differences between the models affect the detailed behaviour in the strongly absorbed regime, the position of the spectral recovery and the overall LIV-induced high-energy behaviour remain essentially unchanged across all tested EBL models. This indicates that, once the recovery feature appears, its energy location is primarily governed by the underlying LIV scale rather than the details of the EBL modelling. Therefore, the main phenomenological conclusion discussed in this work, namely the relation between the recovery energy and the underlying LIV scale, appears robust with respect to  variations across different EBL models.

The LHAASO collaboration used their GRB~221009A observations to rescale the Saldana-Lopez model (see Fig. 4 of~\cite{LHAASA:2023pay}), obtaining lower intensity in the $\lambda > 28\, \mu\mathrm{m}$ range while assuming standard Lorentz-invariant physics. We included the LHAASO rescaled model in our comparison to demonstrate that the resulting recovery peaks remain consistent with those obtained using the other EBL models considered in this work. However, it should be noted that the effect of lower EBL intensity is reduced gamma-ray absorption, the same effect that subluminal LIV produces. Interestingly, the most recent EBL models by Saldana-Lopez et al. (2021) and Finke et al. (2022) are relatively similar, while the LHAASO-rescaled EBL model is substantially lower in the $\lambda > 28\, \mu\mathrm{m}$ range.

\begin{center}
\includegraphics[width=\linewidth]{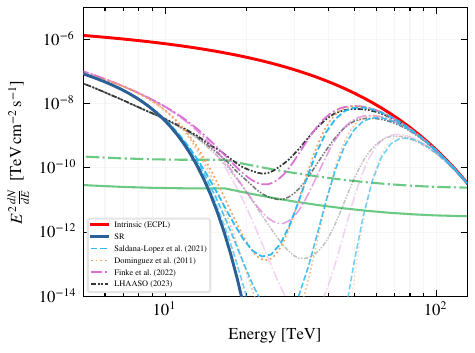}
\captionof{figure}{Intrinsic and absorbed (SR and LIV) BOAT gamma-ray spectra obtained using the Saldana-Lopez EBL model (2021)~\cite{Saldana-Lopez:2020qzx} (blue solid and dashed curves; see~\autoref{fig:boat_liv}), together with the corresponding LIV spectra computed using the Dom\'inguez (2011)~\cite{Dominguez:2010bv} (orange dotted curves), Finke (2022)~\cite{Finke:2022uvv} (pink dash-dotted curves) and the LHAASO-rescaled Saldana-Lopez model (2023)~\cite{LHAASA:2023pay} (black dash-double-dotted curves) EBL models. Within each color scheme, the darkest, lighter, and faintest curves correspond in order to the following choices of the LIV scale: $\Lambda = [1.0,\,1.3,\,2.0]\times10^{-7}\mathrm{E}_{pl}$.}
\label{fig:boatdominguez} 
\end{center}

\section{Influence of alternative pair-production cross section prescriptions}

In addition to testing the robustness of our results against different EBL models, we also investigated the dependence of the predicted SEDs on the choice of LIV-modified $\gamma\gamma\to e^-e^+$ cross section.

The results presented in the main text use the first-principles calculation derived in~\cite{Carmona:2024thn}. Prior to this result, LIV studies relied on approximate prescriptions for the pair-production cross section. Two commonly used approaches, both based on the standard Breit--Wheeler expression, were: \\
(i) introducing an ad hoc cutoff at the second threshold (BW with cutoff), and \\
(ii) replacing the standard Mandelstam variable $s$ with the modified quantity $s' = s - E^4/\Lambda^2$ (BW with modified $s$).

In~\autoref{fig:cross_sections}, we compare the absorbed spectra obtained using these two approximate prescriptions with the spectra derived from the exact calculation. As anticipated in~\cite{Carmona:2024thn}, both approximate expressions systematically overestimate the interaction probability across the physical kinematical range. Consequently, they predict a lower transparency of the Universe and shift the LIV-induced spectral recovery toward higher gamma-ray energies for the same value of the LIV scale $\Lambda$.

Phenomenologically, this implies that using the approximate cross sections leads to lower inferred values of the LIV scale. For example, the spectral recovery obtained using the exact calculation with $\Lambda = 1.3\times10^{-7}E_{\mathrm{Pl}}$ occurs at approximately the same energy as that obtained using the approximate prescriptions with $\Lambda = 1.0\times10^{-7}E_{\mathrm{Pl}}$.

\begin{center}
\includegraphics[width=\linewidth]{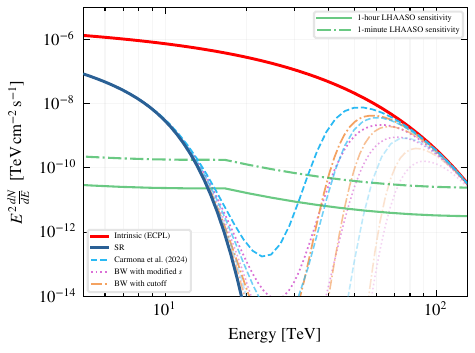}
\captionof{figure}{Comparison of the absorbed BOAT spectra obtained using the first-principles calculation of the LIV-modified BW cross section from Carmona et al.~(2024)~\cite{Carmona:2024thn} and two approximate prescriptions previously used in the literature: the BW cross section with a cutoff at the second threshold (orange dash-dotted curves) and BW with modified-$s$ prescription (pink dotted curves). Within each color scheme, the darkest, lighter, and faintest curves correspond in order to the following choices of the LIV scale: $\Lambda = [1.0,\,1.3,\,2.0]\times10^{-7}\mathrm{E}_{pl}$.}
\label{fig:cross_sections}
\end{center}

\section{LIV signatures from AGN observations with future gamma-ray detectors}

Active galactic nuclei (AGNs) satisfy the condition of having hard and bright spectra, making them good candidates for LIV studies, besides GRBs. Blazar (jetted AGN with jet pointing at our line of sight) PKS 2155-304 ($z=0.116$) is a south sky source that has been observed by both H.E.S.S. and MAGIC in the VHE gamma-ray range~\cite{Aharonian:2007ig,MAGIC:2012iui}. Because of its position, it is a good candidate for LIV studies with future-generation facilities located in the southern hemisphere such as CTAO South and SWGO.

In~\autoref{fig:pks}, we show an observational prospect for the LIV-induced spectrum of PKS 2155-304 considering attenuation by the EBL~\cite{Saldana-Lopez:2020qzx} and CMB. The intrinsic spectrum is taken from the \textit{Fermi}-LAT catalogue (4FGL-DR4)~\cite{Fermi-LAT:2022byn}, which corresponds to an extrapolation to observations of steady emissions in the $50\,\text{MeV}-1\,\text{TeV}$ energy range, modeled with a LogParabola (LP). However, this extrapolation is unrealistic for multi-TeV energies. Following~\cite{CTAO_AGN_inprep} we add an exponential cutoff (ECLP) at 10~TeV that accounts for possible suppression of emissions at higher energies caused by the intrinsic spectrum curvature and/or Klein-Nishina effects. Given SWGO 1 year, and CTAO South 50 hours estimated sensitivities, we conclude that the estimated intrinsic flux of PKS~2155-304 is not above detection threshold for LIV-induced effect for $\Lambda \gtrsim 10^{-7}\,\mathrm{E}_{pl}$. Nevertheless, states of increasing flux, frequent for this source \cite{HESS:2004pzz} and AGNs in general, place this class of targets among the suitable candidates to explore the quadratic subluminal LIV parameter space.

\begin{center}
\includegraphics[width=\linewidth]{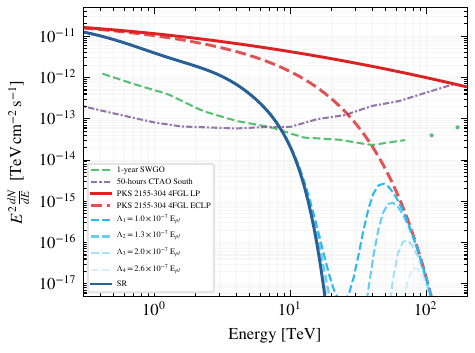}
\captionof{figure}{LIV observational prospect for the blazar PKS~2155-304 in the subluminal quadratic scenario. The red solid line shows the intrinsic spectrum taken from 4FGL catalogue modeled as LP, meanwhile the red dashed line shows the 4FGL spectrum with exponential cut-off where $E_{cut}=10$ TeV. The solid blue line line shows absorbed spectrum in SR scenario assuming EBL+CMB background, while dashed light-blue curves show absorbed spectrum assuming also different scales of LIV. The EBL model used is that of Saldana-Lopez~\cite{Saldana-Lopez:2020qzx}. The green curve (circles) correspond to the SWGO sensitivity (estimated sensitivity extrapolation) for an observation time of 1 year taken from~\cite{SWGO:2025taj}, and the purple curve represents the CTAO South sensitivity obtained from~\cite{CTAO_performance} for an observational period of 50 hours.}
\label{fig:pks} 
\end{center}

\bibliographystyle{cas-model2-names}

\bibliography{biblio}

\end{document}